\documentclass[12pt]{iopart}
\usepackage{iopams}
\usepackage{graphicx}
\usepackage{amstext}
\usepackage{hyperref}
\usepackage{cleveref}
\usepackage{cite}
\usepackage[normalsize]{subfigure}

\usepackage{color}
\usepackage[utf8]{inputenc} 
\usepackage[T1]{fontenc} 
\usepackage[english]{babel} 
\renewcommand{\vec}{\mathbf}

\newcommand{\tauMSD}{\ensuremath{\tau_{\mathrm{MSD}}}}
\newcommand{\tauEE}{\ensuremath{\tau_{\mathrm{EE}}}}

\newcommand{\tauLJ}{\tau_\mathrm{LJ}}
\newcommand{\Dp}{\ensuremath{D_\mathrm{p}}}
\newcommand{\Ds}{\ensuremath{D_\mathrm{s}}}
\newcommand{\Dl}{\ensuremath{D_\mathrm{large}}}
\newcommand{\rcm}{\ensuremath{\vec{r}_\mathrm{cm}}}
\newcommand{\kB}{k_{\mathrm{B}}}
\newcommand{\Np}{N_{\mathrm{p}}}
\newcommand{\Nc}{N_{\mathrm{c}}}

\newcommand{\UFENE}{U_\mathrm{FENE}}
\newcommand{\eab}{\epsilon_{\alpha\beta}}
\newcommand{\epp}{\epsilon_\mathrm{pp}}
\newcommand{\esp}{\epsilon_\mathrm{sp}}

\newcommand{\ess}{\epsilon_\mathrm{ss}}
\newcommand{\reei}{\mathbf{r}^\mathrm{ee}_i}
\newcommand{\sigmaab}{\sigma_{\alpha\beta}}
\newcommand{\sigmapp}{\sigma_\mathrm{pp}}
\newcommand{\sigmass}{\sigma_\mathrm{ss}}
\newcommand{\sigmasp}{\sigma_\mathrm{sp}}

\newcommand{\rab}{r_{\alpha\beta}}
\newcommand{\rc}{r_\mathrm{c,\alpha\beta}}

\newcommand{\ULJab}{U^\mathrm{LJ}_{\alpha\beta}}

\begin{document}

\title[]{Multiple character of non-monotonic size-dependence for relaxation dynamics in polymer-particle and binary mixtures}

\author{Elias M. Zirdehi$^1$, Thomas Voigtmann$^{2,3}$, Fathollah Varnik$^1$}

\address{$^1$ Interdisciplinary Centre for Advanced Materials Simulation (ICAMS), Ruhr-Universität Bochum, Universitätsstraße 150, 44801 Bochum, Germany}

\address{$^2$ Department of Physics, Heinrich-Heine Universität Düsseldorf, Universitätsstr. 1, 40225 Düsseldorf, Germany}

\address{$^3$ Institut für Materialphysik im Weltraum, Deutsches Zentrum für Luft- und Raumfahrt (DLR), 51170 Köln, Germany}

\ead{fathollah.varnik@rub.de}

\begin{abstract}
Adding plasticizers is a well-known procedure to reduce the glass transition temperature in polymers. It has been recently shown that this effect shows a non-monotonic dependence on the size of additive molecules~[The Journal of Chemical Physics 150 (2019) 024903]. In this work, we demonstrate that, as the size of the additive molecules is changed at fixed concentration, multiple extrema emerge in the dependence of the system's relaxation time on the size ratio. The effect occurs on all relevant length scales including single monomer dynamics, decay of Rouse modes and relaxation of the chain's end-to-end vector. A qualitatively similar trend is found within mode-coupling theoretical results for a binary hard-sphere (HS) mixture. An interpretation of the effect in terms of local packing efficiency and coupling between the dynamics of minority and majority species is provided.
\end{abstract}


\section{Introduction}

To identify the physics underlying the glass formation in amorphous materials has been one of the challenging problems in the field of soft condensed matter physics over the last several decades, and still demands further studies~\cite{Binder1999b}. During the glass transition in a material, a sharp increase of relaxation times occurs in a narrow temperature interval, while the static glassy structure undergoes no distinctive change. This feature has been investigated by a myriad of works in experiments, simulations and theoretical studies~\cite{Bengtzelius1984,Schweizer1995,Pusey1986,Pusey1987,Goetze1995,Goetze1999,Pham2002}.

An interesting feature of some glass forming systems is the emergence of a non-monotonic dependence of the relaxation dynamics (see~\cite{Varnik2016} and references therein) by changing only one control parameter. For instance, the mode-coupling theory revealed a reentrant glass-transition line in colloidal systems with short-ranged attractive interactions~\cite{Dawson2000}.
This prediction was later confirmed by experiments on colloid-polymer mixtures~\cite{Eckert2002,Poon2002,Poon2003,Pham2004,Charbonneau2007,Hendricks2015} as well as computer simulations~\cite{Pham2002}. There have been also other studies in the literature with regard to the reentrant glass transition with different mechanisms such as quantum effects~\cite{Markland2011} and confinement~\cite{Lang2012,Lang2013,Mandal2014}.

In recent works~\cite{Hendricks2015,Lazaro2019}, distinct scenarios are reported for the glass transition in  asymmetric binary (colloidal) mixtures. In addition to the possibility of gel formation~\cite{Hendricks2015}, the larger component can undergo a glass transition while the smaller one remains ergodic. Another path is marked by simultaneous loss of ergodicity in both components and is therefore termed as a double glass transition. While the latter is marked by a strong coupling between the dynamics of the two species, the single glass state is ascribed to a decoupling of self-dynamics from collective dynamics at short length scales~\cite{Hendricks2015,Lazaro2019}.

The above studies explore the available states in binary mixtures by varying the packing fraction at a fixed size ratio. There have been few studies focusing on the effects of size ratio in bi-disperse systems of fixed composition~\cite{Moreno2006,Zirdehi2019}. A central result of these studies is the observation of a non-monotonic dependence of the glass transition on the size-ratio of the constituents. The systems investigated include binary mixture of soft spheres~\cite{Moreno2006} and polymer-additive blends~\cite{Zirdehi2019}.

A non-monotonic size-dependence of glass transition is also predicted by a mode-coupling theory (MCT) for binary hard-sphere mixtures~\cite{Goetze2003,Voigtmann2011} and by generalizations of nonlinear Langevin equations~\cite{JuarezMaldonado2008,Zhang2018}. A detailed survey of MCT calculations reveals the existence of multiple extrema in relaxation time versus concentration~\cite{Voigtmann2011}. This means that, at a fixed size ratio, varying concentration of the minority component leads to multiple passages between the liquid-like and glassy states (liquid-glass-liquid-glass-...). A question of interest here is whether this multiple re-entrance can occur at a constant concentration but by varying the size-ratio of the two components, and whether it can rationalize the non-monotonic effect that was recently found for the plasticization of polymers by addition of small molecules \cite{Zirdehi2019}.

In the current work, we explore this issue by combining MD simulations of a polymer/additive system and MCT analysis of a colloidal hard sphere mixture. In both cases, the concentration of additive/small component is kept constant but the size-ratio small/large particles is varied. It will be shown that, indeed, additional extrema exist in the dependence of structural relaxation time on size-ratio. The fact that qualitatively this phenomenon occurs both in entangled polymers containing small spherical additives and in binary mixture of colloidal articles calls for an interpretation in terms of local competing packing effects. A discussion of this issue is provided. 

\section{Model}

\subsection{Polymer-additive simulation model}

The polymer is modeled as a linear chain of spherical particles connected via a finite extensible nonlinear elastic (FENE) force. The corresponding interaction potential reads~\cite{Kremer1988,Baschnagel2005},
\begin{equation}
\UFENE(r)=-\frac{1}{2}kR_{0}^2\ln \Big[1-\Big(\frac{r}{R_0}\Big) ^{2}\Big],
\label{Eq:Model-FENE}
\end{equation}
where \(k=30\epp/\sigmapp^2=30\) is the strength factor and \(R_0=1.5\) the breaking limit of covalent bonds. Throughout this paper, the index p refers to polymeric beads (monomers) and s to (small) spherical particles added to the system. In addition to the FENE-potential which acts only between adjacent monomers along the chains' backbone, all particle pairs interact via a Lennard-Jones (LJ) potential,
\begin{equation}
\ULJab(\rab)=4 \eab \Big[\Big(\frac{\sigmaab}{\rab}\Big)^{12}-\Big(\frac{\sigmaab}{\rab}\Big)^{6}\Big].
\label{Eq:Model-LJ}
\end{equation}
In Eq.~(\ref{Eq:Model-LJ}), $\alpha, \beta \in \{\text{p}, \text{s}\}$ and $\rab$ is a short hand notation for the distance between a particle, \(i\), of type \(\alpha\) and another one, \(j\), of type \(\beta\):  \(\rab= |\vec r_{i,\alpha} -\vec r_{j,\beta} |  \). The LJ potential is truncated at a cutoff radius of \( \rc=2\times2^{1/6}\sigmaab \). The monomer diameter, \(\sigmapp\), is kept constant throughout the simulation and defines the unit of length (a convenient way to achieve this is to set \(\sigmapp\equiv 1\)). The size disparity is defined via $\delta=\sigmass/\sigmapp$. The parameter \(\sigmasp\) is chosen as the arithmetic mean, \(\sigmasp=0.5(\sigmapp+\sigmass)\). For simplicity, the energy scale of the LJ potential is set to unity regardless of the particle type, i.e. \(\epp=\ess=\esp=1\).  The mass of an additive particle is set to be equal to that of a monomer, $m_\text{s}=m_\text{p}=1$.

Temperature is measured in units of $ \epp/\kB $ with the Boltzmann constant $\kB$. All other quantities are given as a combination of the above described units. The unit of time, for example, is given by $\tauLJ=(m\sigmapp^2/\epp)^{1/2}$ and that of pressure is \( \epp/\sigmapp^3 \). All quantities addressed here are expressed in this set of reduced LJ units. The equations of motion are integrated using the Velocity-Verlet algorithm with a time step of $\delta t=0.003$. All the simulations are performed by the open source molecular dynamics simulator LAMMPS~\cite{Plimpton1995}.

The number of monomers per chain is $\Np=10$. This choice is motivated by the need to equilibrate the system at all investigated temperatures, while keeping the computational cost acceptable. The total number of particles is $N=4000$.
Simulations are first prepared in the $NpT$-ensemble at a pressure of $p=1$. We then switch to the $NVT$-ensemble for dynamics measurements. This way, we avoid undesirable effects on dynamics which could originate from the fluctuations of the simulation cell size~\cite{Varnik2002c}. The number-concentration of additive molecules is kept constant at $20\%$. In contrast to this, the size disparity is varied from 0.3 to 1 by adjusting the size of the additive particles, $\sigmass$.

\subsection{MCT for binary colloid mixtures}

The central quantity for MCT is the matrix of time-dependent density autocorrelation
functions,
$\Phi_{\alpha\beta}(q,t)=\langle\varrho_\alpha(\vec q,t)^*\varrho_\beta(\vec q,0)\rangle$,
where $\varrho_\alpha(\vec q,t)=\sum_{k=1}^{N_\alpha}\exp[i\vec q\cdot\vec r_{\alpha,k}(t)]$ are the microscopic density fluctuations to wave vector $\vec q$
associated to the $N_\alpha$ particles of type $\alpha$. Its equal-time
limit is just the static structure factor, $\boldsymbol\Phi(q,0)=\boldsymbol S(q)$.
Following a
projection-operator formalism \cite{Goetze2009}, one derives a time-evolution
equation of the form (in matrix notation)
\begin{equation}\label{eq:morizwanzig}
  \boldsymbol J(q)^{-1}\cdot\partial_t^2\boldsymbol\Phi(q,t)
  +\boldsymbol S(q)^{-1}\cdot\boldsymbol\Phi(q,t)
  +\int_0^t\boldsymbol M(q,t-t')\cdot\partial_{t'}\boldsymbol\Phi(q,t')\,dt'
  =\boldsymbol0\,.
\end{equation}
The $J_{\alpha\beta}(q)=q^2(k_BT/m_\alpha)x_\alpha\delta_{\alpha\beta}$
set the time scales for the short-time motion, chosen here to fix the unit of time through the thermal velocity of the bigger particles.
For simplicity, we set both masses equal (as in the simulation); the
long-time dynamics predicted by MCT does not crucially depend on this
choice as long as the mass ratio is not too extreme \cite{Mandal2018}.
The memory kernel $\boldsymbol M(q,t)$ describes the slow structural-relaxation dynamics, and is approximated by MCT as a bilinear functional of the density-correlation functions \cite{Goetze2009,Goetze2003},
\begin{equation}\label{eq:mctmemory}
  M_{\alpha\beta}(q,t)=\frac{N}{2Vq^2x_\alpha x_\beta}\int\frac{d^3k}{(2\pi)^3}
  V_{\alpha\gamma\lambda}(\vec q,\vec k)\Phi_{\gamma\delta}(k,t)
  \Phi_{\lambda\mu}(p,t)V_{\beta\delta\mu}(\vec q,\vec k)\,,
\end{equation}
where $p=|\vec q-\vec k|$, and with vertices that are determined entirely
by the static structure of the system. In a common approximation that
simplifies triplet correlations,
\begin{equation}\label{eq:mctvertex}
  V_{\alpha\gamma\lambda}(\vec q,\vec k)
  =\delta_{\alpha\lambda}(\vec q\cdot\vec k)c_{\alpha\gamma}(k)/q
  +\delta_{\alpha\gamma}(\vec q\cdot\vec p)c_{\alpha\lambda}(p)/q\,.
\end{equation}
The static structure factors are assumed to be known; in the present work,
we calculate them from the Percus-Yevick approximation for hard-sphere
mixtures~\cite{Hansen1990}, to obtain a closed fully analytic theory.

The long-time limits of the density correlation functions,
$\boldsymbol F(q)=\lim_{t\to\infty}\boldsymbol\Phi(q,t)$ are called the glass form factors.
They are identically zero in the liquid, and non-vanishing in the ideal
glass.
From Eqs.~(\ref{eq:morizwanzig}) and (\ref{eq:mctmemory}) one obtains
an algebraic equation for the $\boldsymbol F(q)$, viz.\
$
(\boldsymbol S(q)-\boldsymbol F(q))^{-1}=\boldsymbol S(q)^{-1}+\boldsymbol M(q)
$,
where $\boldsymbol M(q)=\boldsymbol M[\boldsymbol F,\boldsymbol F;q]$ is the MCT memory kernel evaluated with the form factors. The bifurcation points of this equation identify the ideal glass-transition points of MCT and can be found via a simple iteration scheme.

The long-time diffusion coefficients $D_\alpha$ of the individual particles of type $\alpha$ are obtained from the respective tagged-particle density correlation functions, $\phi^s_\alpha(q,t)=\langle\varrho^s_\alpha(\vec q,t)^*
\varrho^s_\alpha(\vec q,0)\rangle$, where $\varrho^s_\alpha(\vec q,t)
=\exp[i\vec q\cdot\vec r^s_\alpha(t)]$ is the single-particle microscopic density fluctuation. In the limit $q\to0$, one obtains an equation for the mean-squared displacement (MSD) of the tagged particle,
\begin{equation}
  \partial_t^2\delta r^2_\alpha(t)+v_{\text{th},\alpha}^2\frac{d}{dt}\int_0^t\hat m^s_\alpha(t-t')
  \partial_{t'}\delta r^2_\alpha(t')\,dt'=6v_{\text{th},\alpha}^2\,,
\end{equation}
with the thermal velocity $v_{\text{th},\alpha}=\sqrt{k_BT/m_\alpha}$.
The analysis of the long-time behavior of this equation yields
\begin{equation}
  D_\alpha=\frac{1}{\int_0^\infty\hat m^s_\alpha(t)\,dt}\,,
\end{equation}
which is finite as long as the tagged-particle memory kernel
$\hat m^s_\alpha(t)$ decays to zero for long times. This is always the case in the liquid, but if the host system is in an ideal-glass state,
weakly coupled tagged particles can still retain a finite diffusivity in this glass.

\section{Particle size effects in polymer-additive systems}

In this section, we analyze the relaxation dynamics in the polymer/additive system for a fixed concentration of the additive component with a focus on the occurrence of multiple extrema as the size of added particles (minority species) is varied.

\subsection{Overall Dynamics}
\label{sec:polymeric-system}

\begin{figure}
	\centering
	(a)\includegraphics[width=0.45\textwidth]{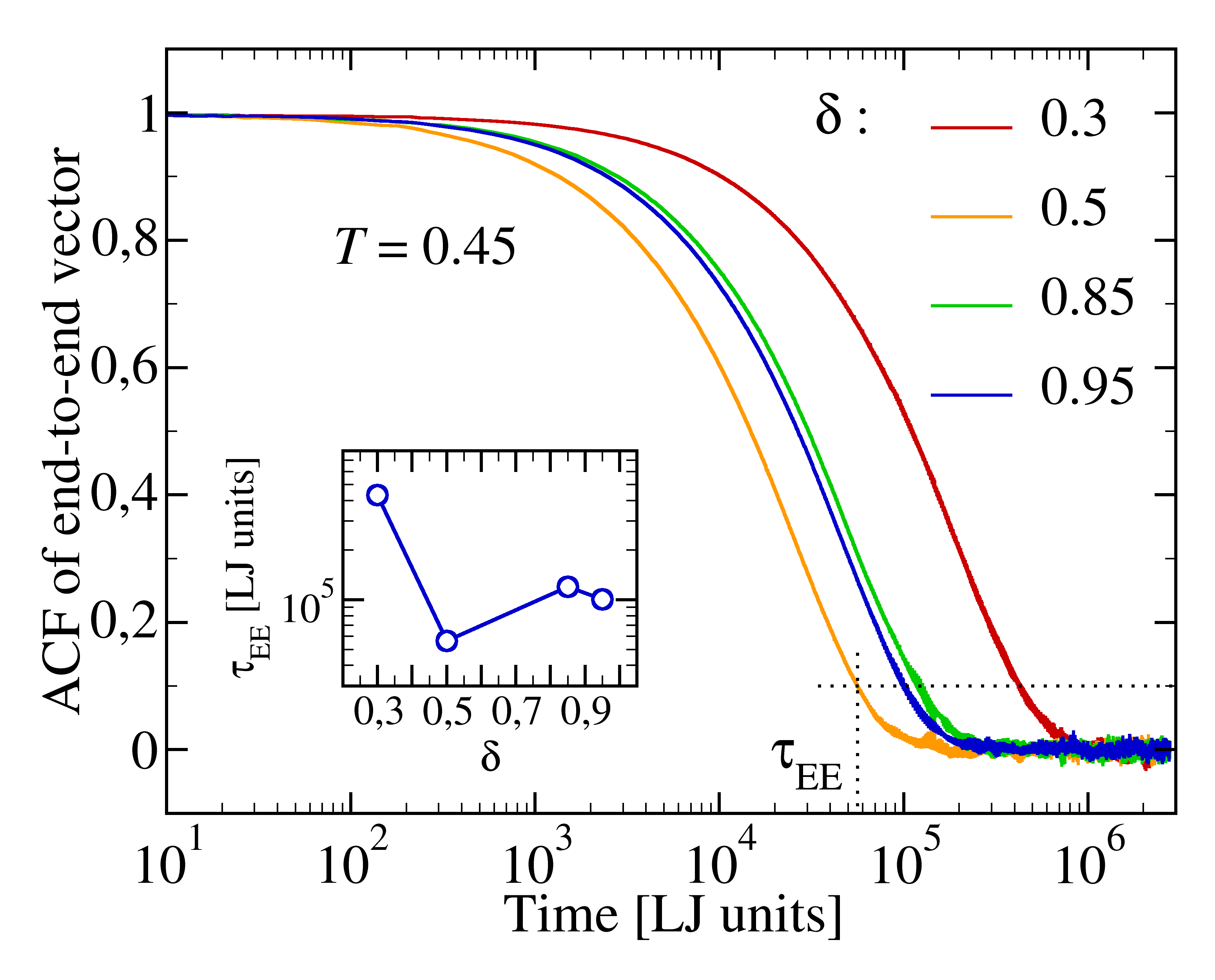}
	(b)\includegraphics[width=0.45\textwidth]{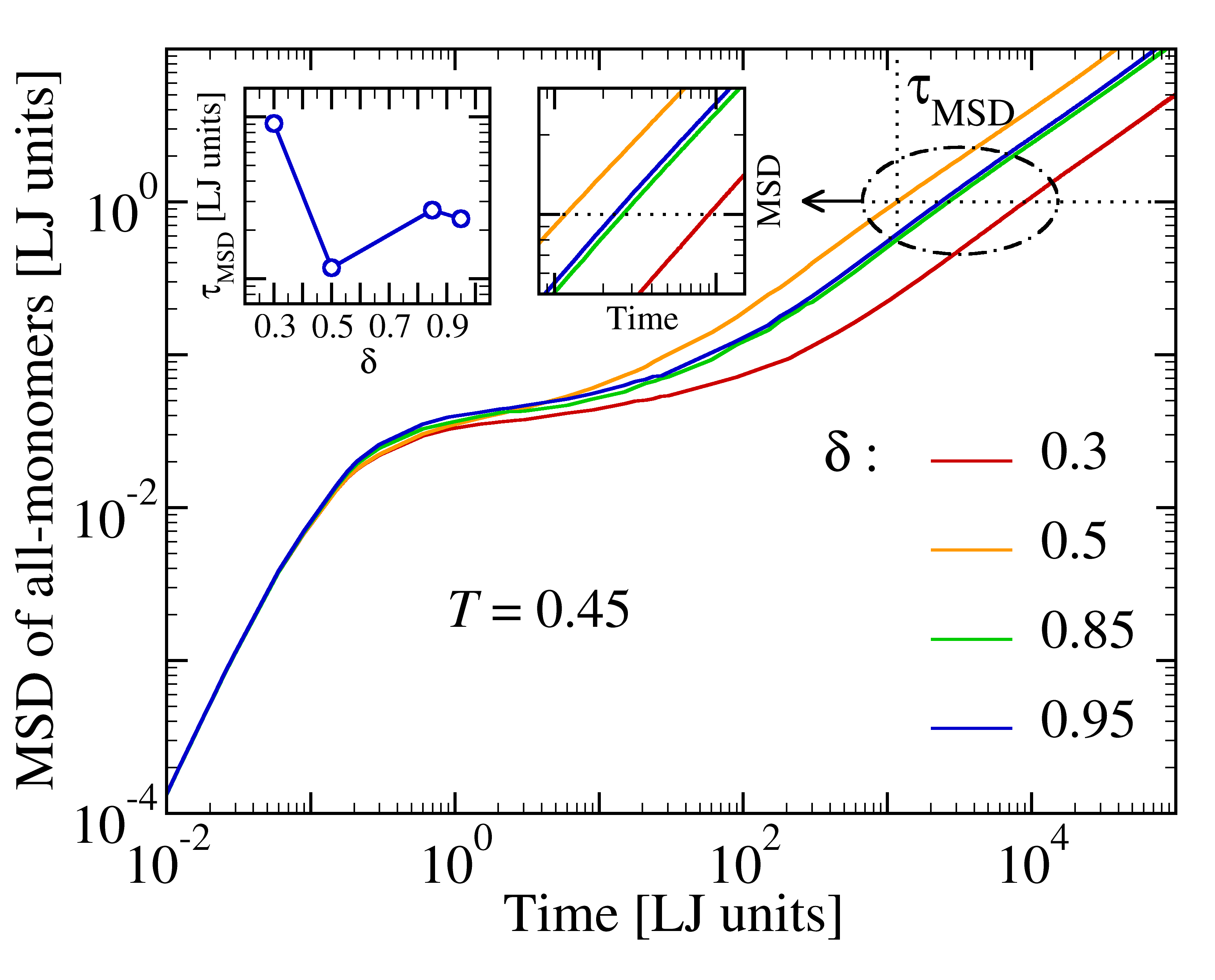}
	\caption{(a) The end-to-end autocorrelation function of polymer chains versus time for the present FENE-polymer model containing a number concentration of 20\% additive molecules. Different curves correspond to different size ratios $\delta=\sigmass/\sigmapp$. The data show a clear signature of a non-monotonic dependence of the relaxation rate on $\delta$. To highlight this observation in a quantitative way, we define a relaxation time, $\tauEE$, as the time needed by the autocorrelation function to decay to a (small) value, here chosen to be 0.1. Dotted lines serve to illustrate this procedure. The thus obtained $\tauEE$ are plotted in the inset versus size disparity and reveal the existence of multiple extrema. (b) An alternative approach to the relaxation dynamics by following the mean-squared displacement of a single monomer. The corresponding relaxation time, $\tauMSD$ is defined as the time during which a monomer travels by its own diameter (see dotted-lines). Importantly, single monomer dynamics and decay of the chain's end-to-end vector depend in qualitatively the same way on $\delta$.}
	\label{fig:EtE+MSD}
\end{figure}

The main observation for the case of the polymer-additive system is illustrated in Fig.~\ref{fig:EtE+MSD}, where both a polymer-specific quantity -- the autocorrelation function of the chain's end-to-end vector -- and a quantity which stands for single particle dynamics -- mean-squared displacement (MSD) of a monomer, averaged over all monomers in the system -- are shown.
The autocorrelation function of the end-to-end vector is obtained from EtE-ACF$(t)=\sum_{i=1}^{\Nc} \left< \reei(t) \cdot \reei(0)\right>/\sum_{i=1}^{\Nc} \left< \reei(0) \cdot \reei(0)\right>$, where $\reei(t)$ denotes the vector connecting the first and the last monomers of the $i$-th chain at time $t$ and the sum runs over all $\Nc$ chains. The all-monomer mean-squared displacement is determined via MSD$(t)=\sum_{i=1}^N \left<[\mathbf{r}_i(t)-\mathbf{r}_i(0)]^2\right>/N$, where the sum runs over all $N$ monomers.

As seen from Fig.~\ref{fig:EtE+MSD}, both the relaxation dynamics of a polymer chain and the single-particle MSD exhibit a non-monotonic dependence in their long-time behavior on the diameter of additive particles. The long-time relaxation first slightly slows down with decreasing the size of the additive from $\delta=1$ to $\delta\approx0.85$; then it speeds up upon lowering the additive size down to $\delta\approx0.5$; and finally, it strongly slows down again at even smaller $\delta$. The dependence of typical relaxation times on $\delta$ thus shows multiple extrema (insets of Fig.~\ref{fig:EtE+MSD}). The fact that both the end-to-end vector ACF and the monomeric MSD show the same qualitative features, suggests that the observed effect is not a consequence of a coupling between chain connectivity and dynamics of additive molecules. Rather, it seems to reflect the influence of added molecules on the local packing of monomers, regardless of their connectivity along the chains' backbone. As will be shown later below, this interpretation is further corroborated by the fact that a binary mixture of hard sphere particles is predicted within MCT to also exhibit qualitatively the same features in terms of size disparity, $\delta$.

\begin{figure}
    \centering
    (a)\includegraphics[width=0.45\textwidth]{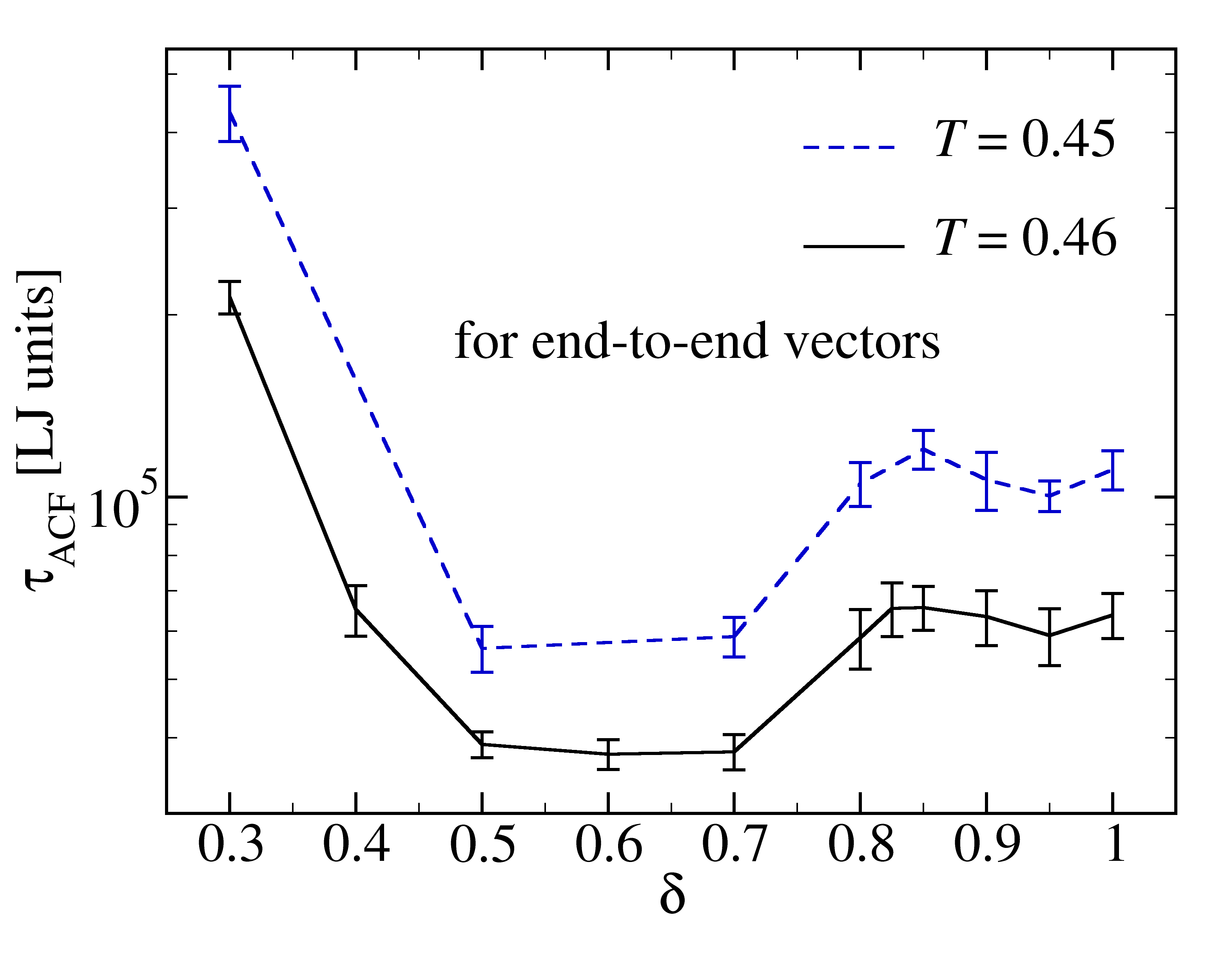}
    (b)\includegraphics[width=0.45\textwidth]{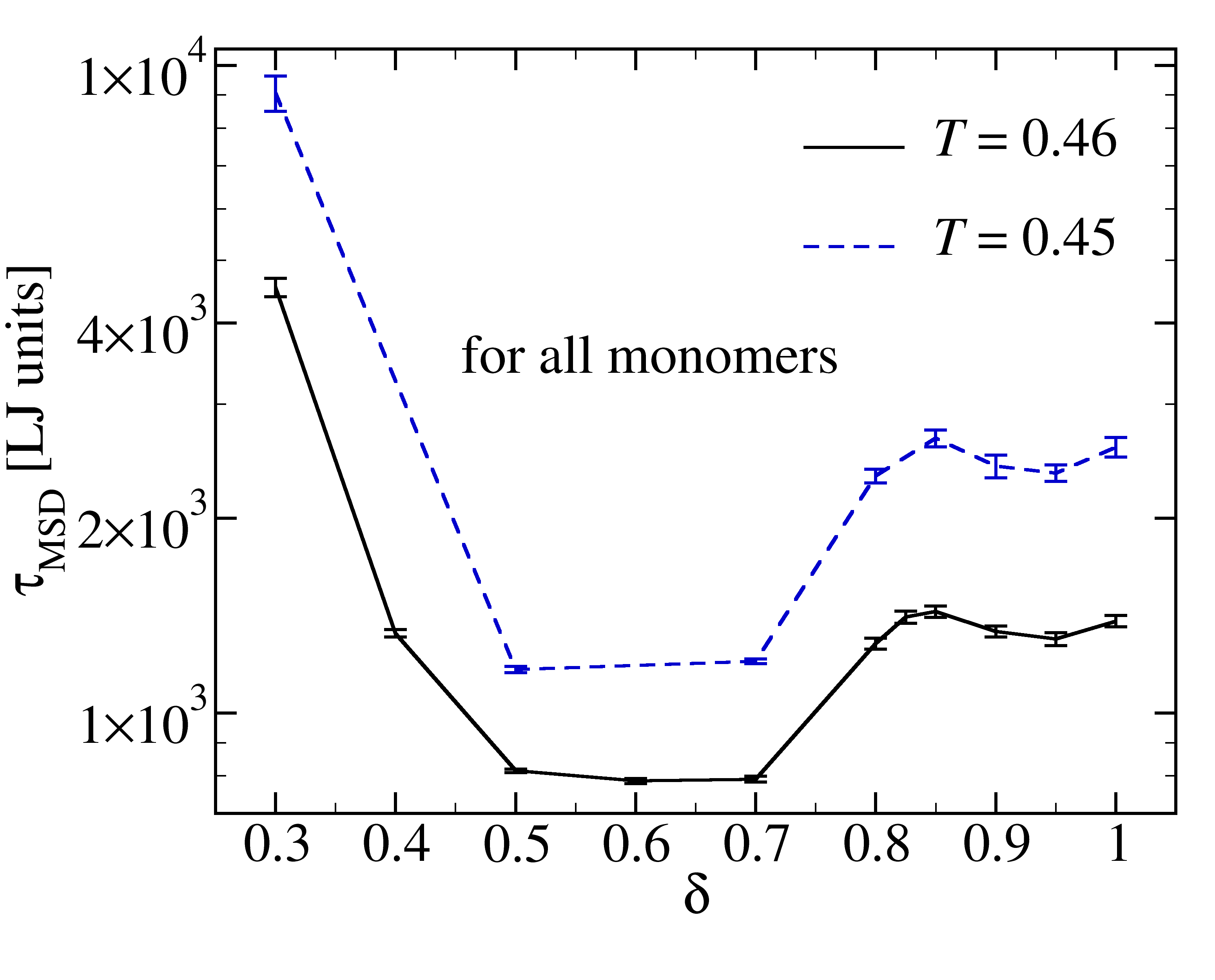}
    \caption{(a) Relaxation times obtained from the decay of auto correlation function of the end-to-end vector for two temperatures from the supercooled state ($T=0.45$ and $T=0.46$). Clearly, the main (global) minimum at a size ratio of $\delta \approx 0.5$~\cite{Zirdehi2019} is followed by a maximum at $\delta \approx 0.85$ and an additional minimum at $\delta\approx 0.95$. Panel (b) shows that the same trend is also observed in the relaxation data extracted from all-monomer mean-squared displacements. Error bars give the 86\% confidence interval.}
   \label{fig:RelaxTime_MSD}
\end{figure}

To provide further quantitative evidence for the existence of multiple extrema, we have performed an extensive set of simulations and have sampled the parameter range for size disparity more densely. Figure~\ref{fig:RelaxTime_MSD} shows the thus obtained results on the structural relaxation times versus $\delta$. The data are extracted  both from the relaxation dynamics of chains' end-to-end vector and from the all-monomer mean-squared displacements. It is noteworthy that, in contrast to the main (global) minimum, which is associated with a relatively large change in the relaxation time and is thus rather easy to detect, resolving the existence of additional extrema requires significant computational effort to reach a sufficiently high signal to noise ratio. As can be inferred from the error bars in the relaxation data, this goal has been achieved convincingly in our simulations.
The effect is demonstrated for two temperatures close to $T_c$ and does not qualitatively depend on temperature in this regime. This suggests that it is an underlying change in the glass-transition point $T_c$ itself as a function of $\delta$ that explains the observed non-monotonic dependence of the relaxation time on the size ratio. In principle, one has to be careful in distinguishing possible pre-asymptotic effects on the dynamics (that may have a different size-ratio dependence) from changes of the glass-transition point \cite{Mandal2018}, but our MCT calculations discussed below support this interpretation.

\begin{figure}
    \centering
    (a)\includegraphics[width=0.45\textwidth]{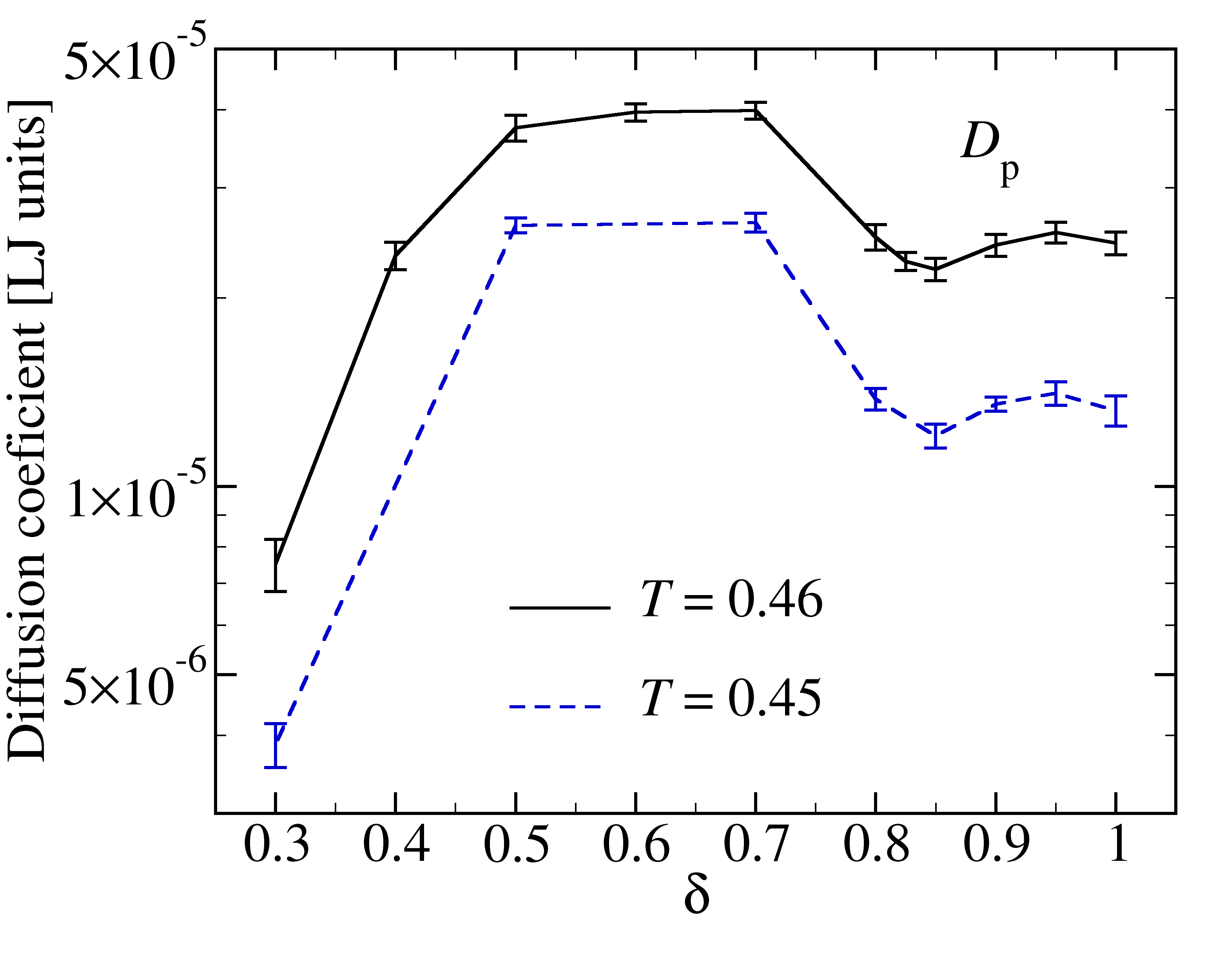}
    (b)\includegraphics[width=0.45\textwidth]{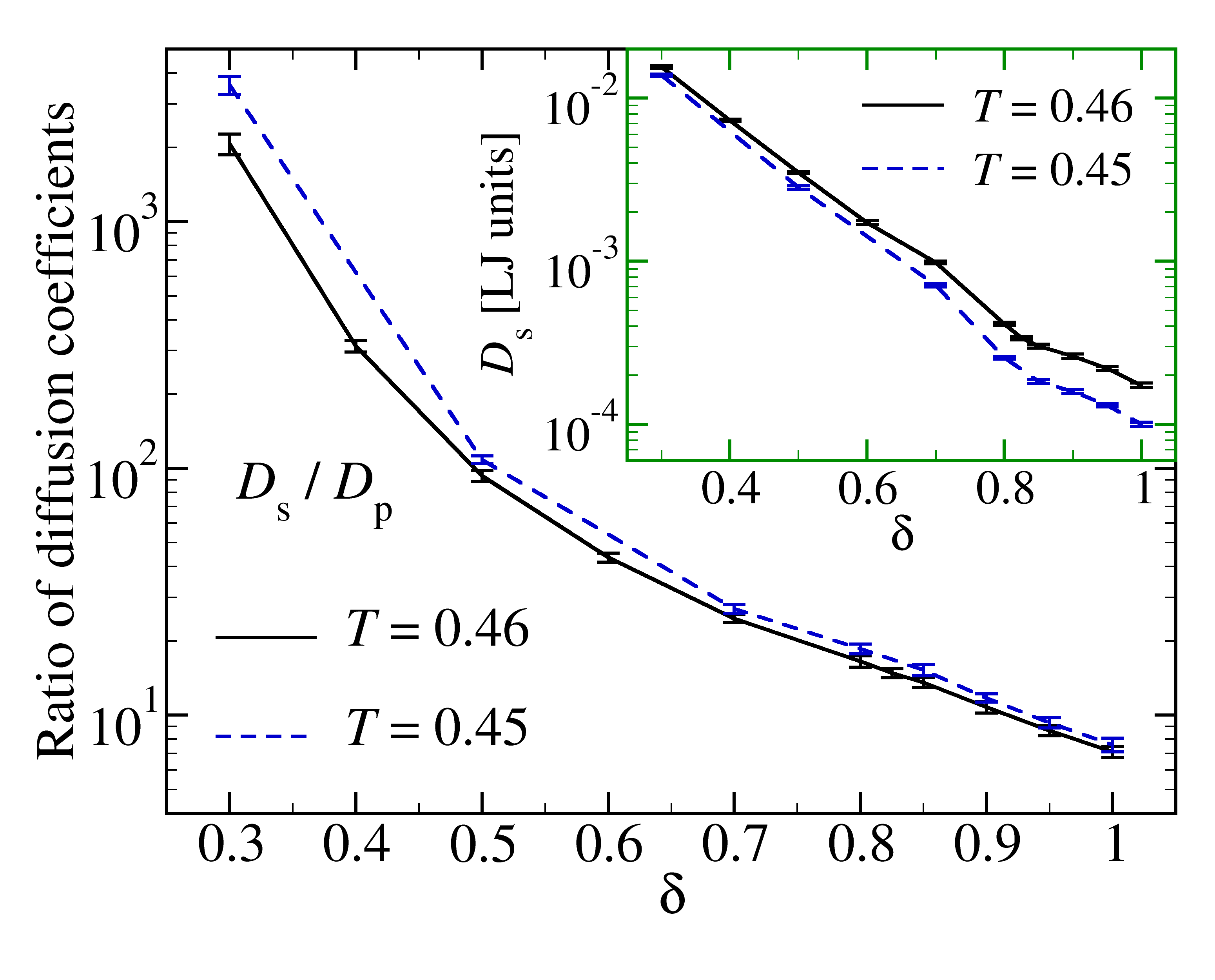}
    \caption{Diffusion coefficient, $\Dp$, of the polymer chains versus the size disparity for two low temperatures in the supercooled regime. $\Dp$ is obtained from the long-time limit of the chains' center of mass MSD. (b) The ratio of single molecules' diffusion coefficients to that of polymer chains, $\Ds/\Dp$ showing a steep increase for $\delta \to 0$. Despite this, the effect of additive molecules on the polymer dynamics weakens as $\delta$ becomes small (see the left panel). The inset shows variation of single molecule's diffusion coefficient, $\Ds$, with the size ratio. Error bars give the 86\% confidence interval.}
    \label{fig:DiffCoef_T46}
\end{figure}

Figure~\ref{fig:DiffCoef_T46}a underlines these findings further by showing polymer diffusion coefficient versus size disparity. Here, diffusion coefficient is extracted from the long time limit of the chains' center of mass displacements, $\Dp=\lim_{t\to \infty} \left< (\vec\rcm(t)-\rcm(0 )^2 \right>/6t$. Noteworthy, as shown in Fig.\ref{fig:DiffCoef_T46}b, the mobility of additive molecules increases rapidly and in a monotonic way as size disparity decreases ($\delta=\sigmass/\sigmapp \to 0$). In contrast to this, the enhancement of polymer dynamics is non-monotonic (Fig.~\ref{fig:RelaxTime_MSD} and Fig.~\ref{fig:DiffCoef_T46}a). A possible way to rationalize this observation is as follows. As the size of additive molecules decreases, they can explore the small available free volume while at the same time experiencing fewer collisions with monomer beads. The coupling between the motion of polymer chains and that of additive molecules thus decreases with smaller size ratios. This competition of increasing mobility and decreasing coupling strength is one of the main reasons for the occurrence of the observed non-monotonic effect~\cite{Zirdehi2019}.

\subsection{Dynamics of Rouse modes}
The above data show that multiple extrema occur both in a polymer specific quantity (decay of the end-to-end vector) and in single particle dynamics (MSD). Here we analyze this issue in more details with a focus on Rouse modes~\cite{Rouse1953}, defined via $\vec X_p(t)=\frac{1}{\Np}\sum_{n=1}^{\Np}\vec r_n(t)\cos(\frac{(n-0.5)p\pi}{\Np}$), $p=0,1,\ldots,\Np-1$~\cite{Varnik2003a}. In this definition, $\vec r_n$ is the position of the $n$-th monomer of a chain. The parameter $p$ plays qualitatively the role of a wave number in Fourier series. Rouse modes thus provide information on chain relaxation dynamics at different length scales, from the entire chain ($p=0$) down to a single monomer ($p=\Np-1$).

\begin{figure}
	\centering
	(a)\includegraphics[width=0.45\textwidth]{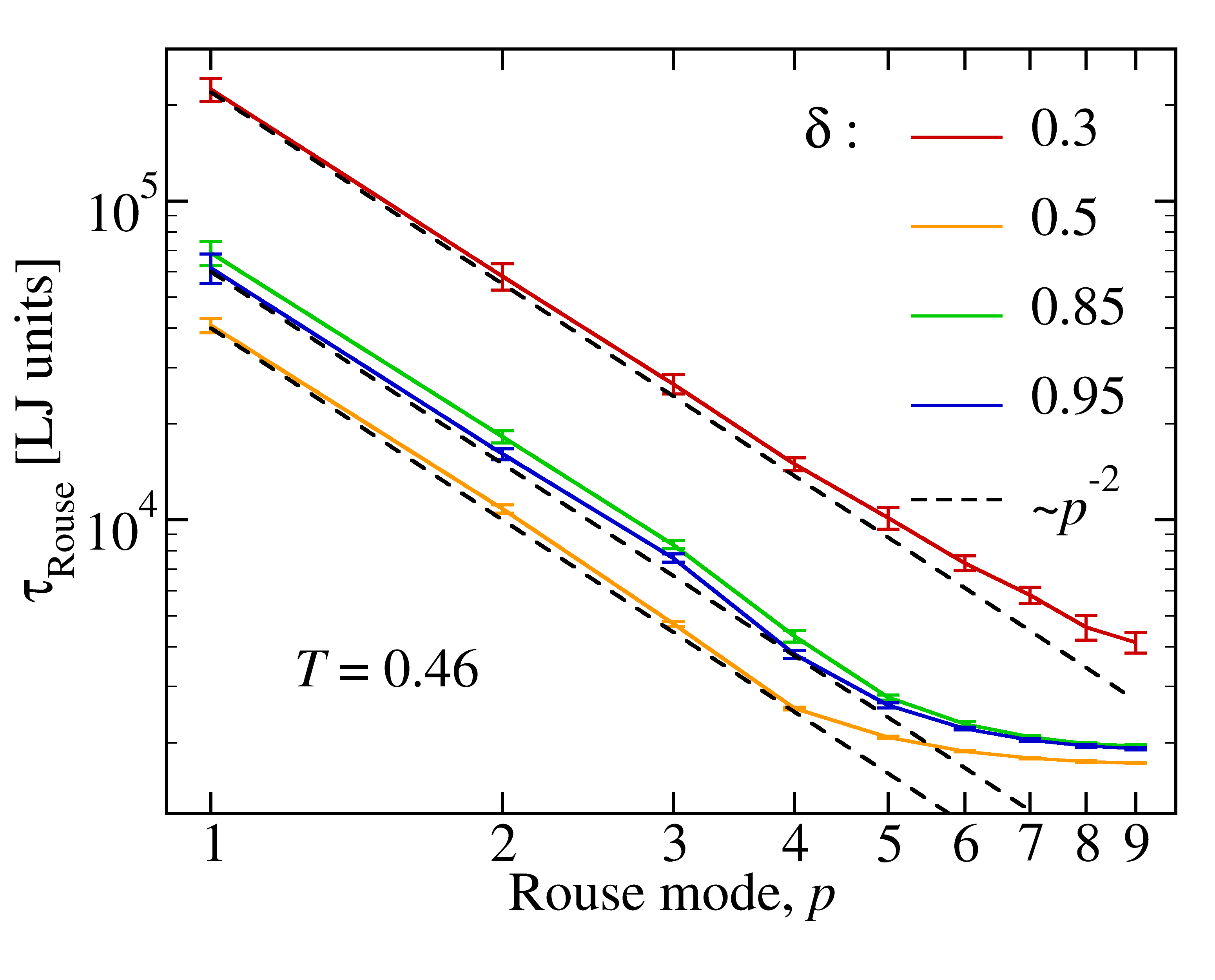}
	(b)\includegraphics[width=0.45\textwidth]{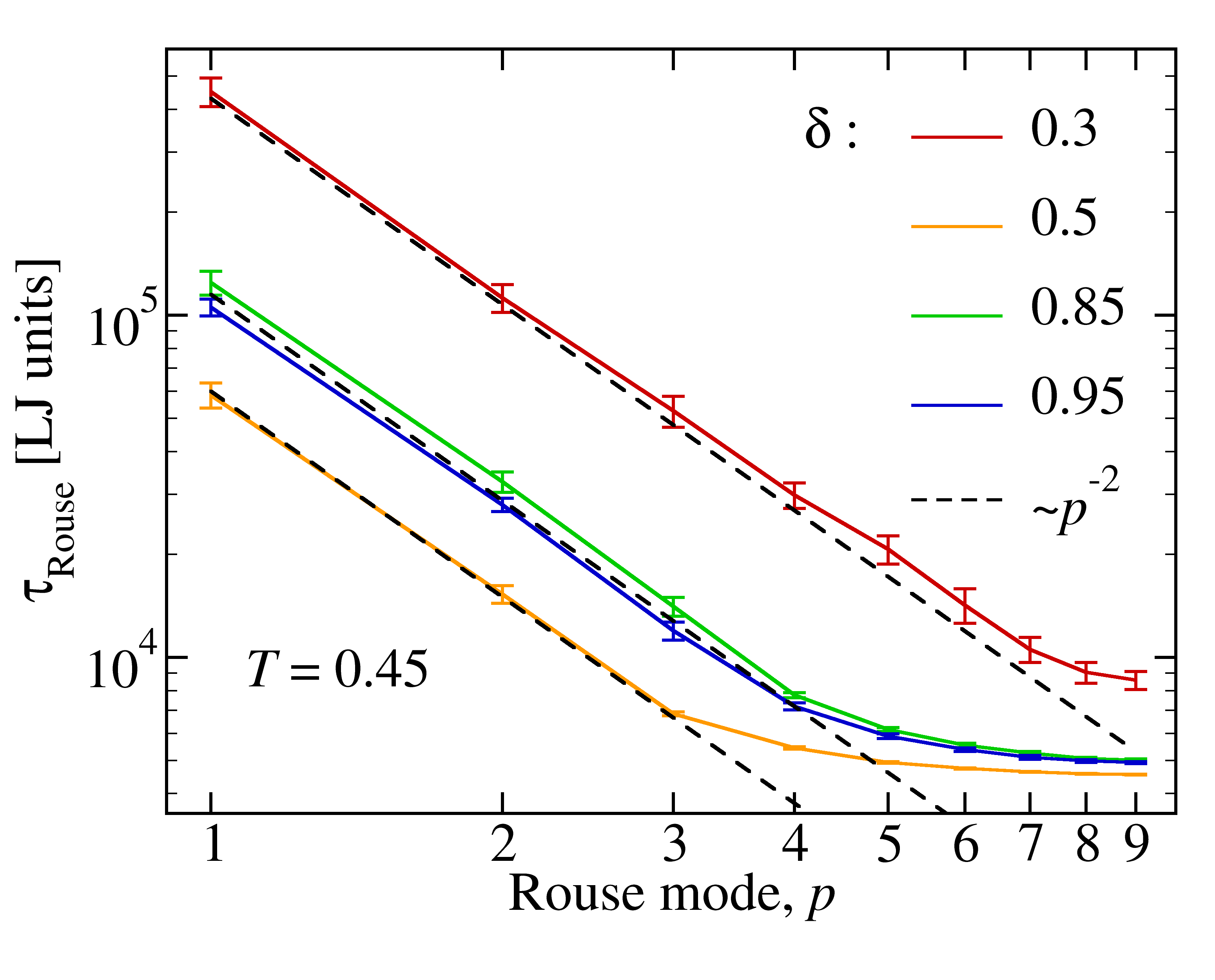}
	\caption{(a) Relaxation time of Rouse modes plotted versus the mode number, $p$, for a polymer melt containing 20\% additive molecules. Different curves correspond to additive-to-monomer size ratios as indicated. The prediction of Rouse model is shown as dashed lines ($\tau_{\mathrm{Rouse}}\propto 1/p^2$)~\cite{Rouse1953}. At small $p$, where Rouse modes sample length scales comparable to extension of a polymer chain, simulation results are in good agreement with the predicted $p^{-2}$-behavior. At smaller length (larger $p$) neglecting volume exclusion makes the predictions of Rouse theory inadequate for the present FENE-polymer. The panel (b) shows the same type of data as in (b) but at a different temperature. The same conclusions can be drawn also from this plot.  Error bars give the 86\% confidence interval.}
	\label{fig:Rouse-tau-vs-p}
\end{figure}

Taking advantage of this flexibility of the Rouse modes in focusing on different length scales, we investigate their relaxation behavior as a function of size ratio. Before doing this, and as a benchmark of our simulations, we first check how the relaxation time of Rouse modes depends on the mode index, $p$. Within the Rouse model, where monomers are subject to random forces acting on them and are at the same time connected via harmonic springs to their neighboring beads, one expects that $\tau_{\mathrm{Rouse}}\propto 1/p^2$~\cite{Rouse1953,Varnik2003a}. As shown in Fig.~\ref{fig:Rouse-tau-vs-p}, for small $p$ (which correspond to large lengths close to the size of a chain), this prediction provides a good approximation to the behavior of Rouse modes within our model system. Deviations from Rouse prediction at large $p$ (small length scales) reflect excluded-volume interactions, which correlate particle motion at short lengths and thus slow down the structural relaxation at these length scales as compared to the predictions of ideal Rouse model, which treats monomers as point particles.

\begin{figure}
	\centering
	(a)\includegraphics[width=0.45\textwidth]{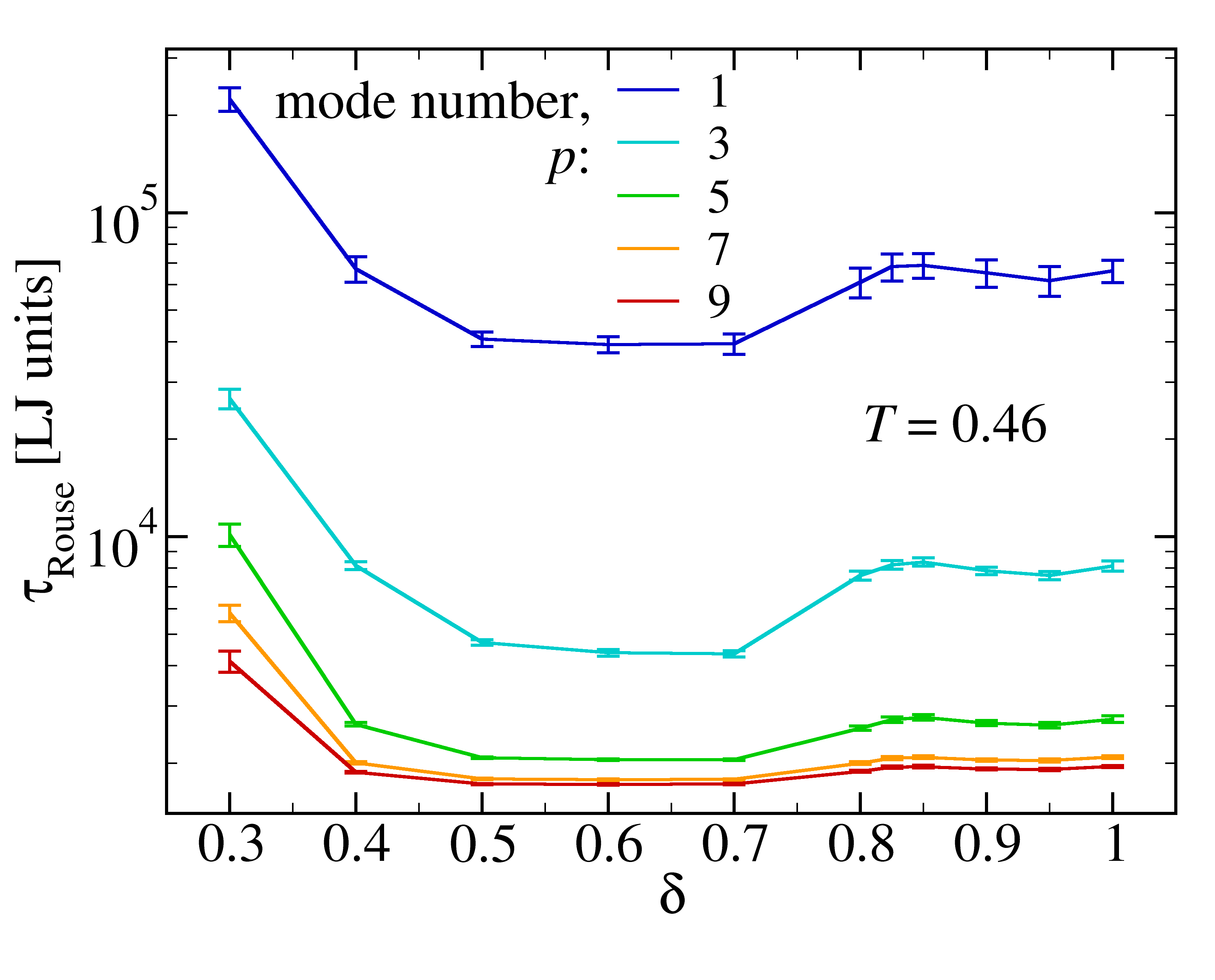}
	(b)\includegraphics[width=0.45\textwidth]{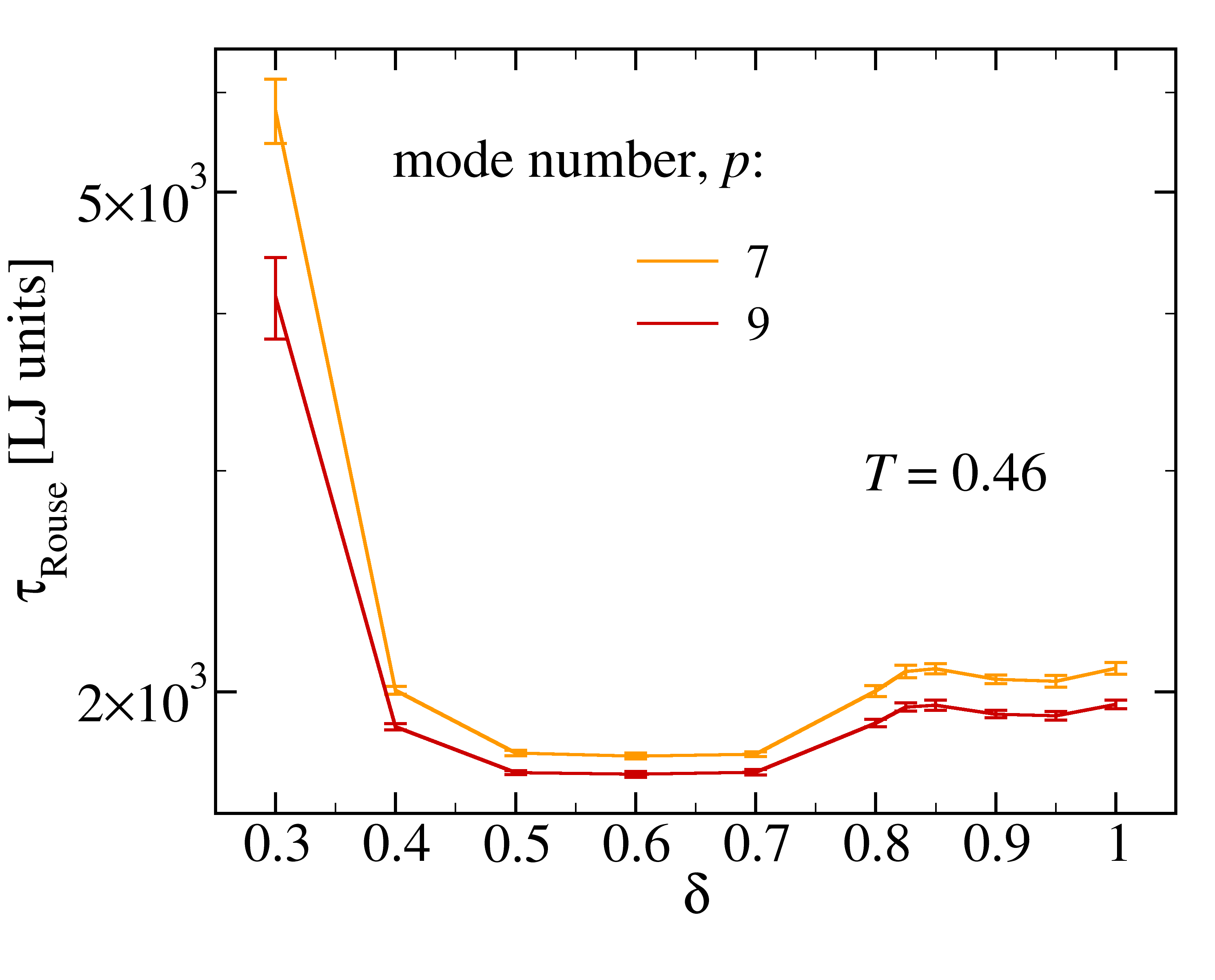}
	(c)\includegraphics[width=0.45\textwidth]{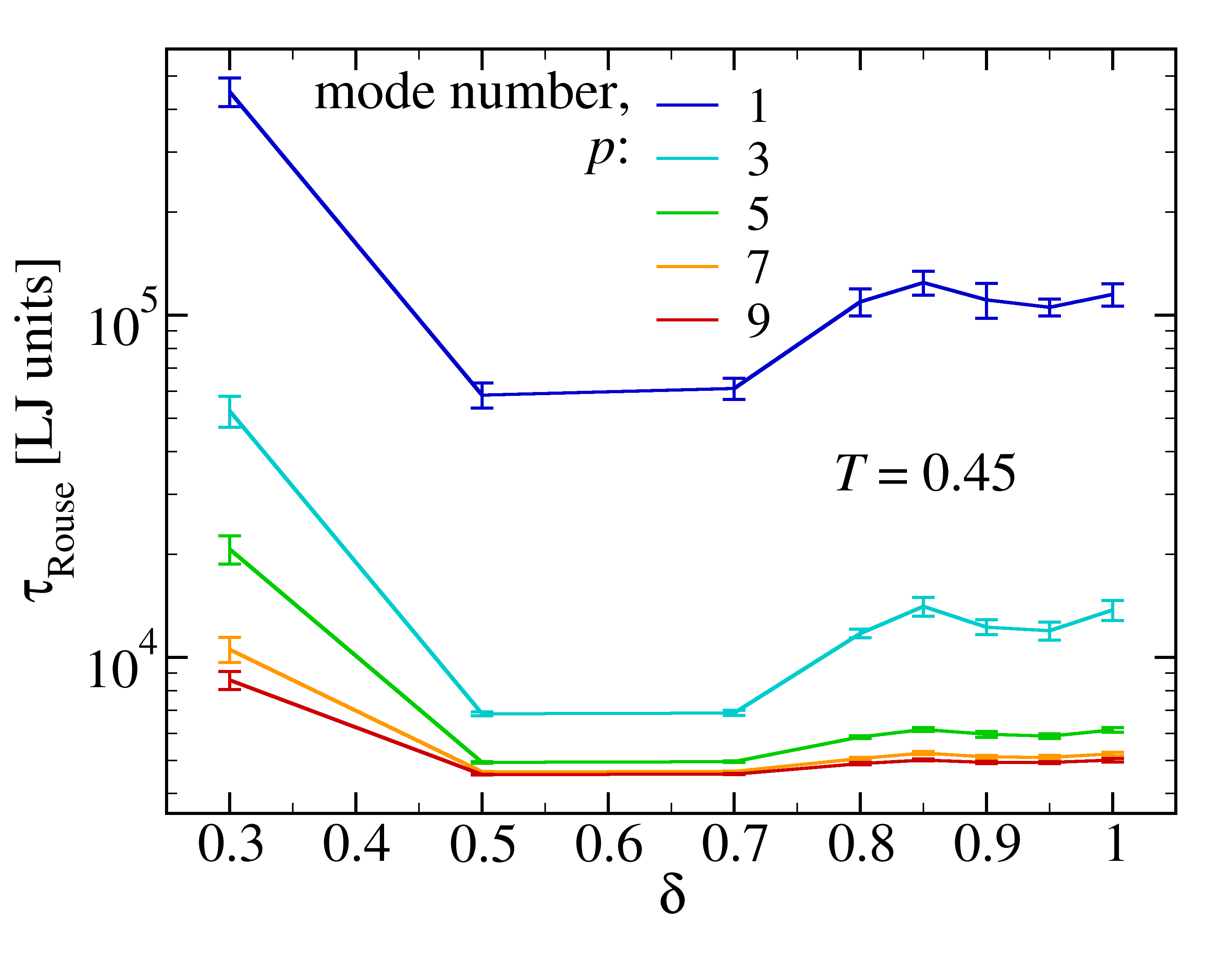}
	(d)\includegraphics[width=0.45\textwidth]{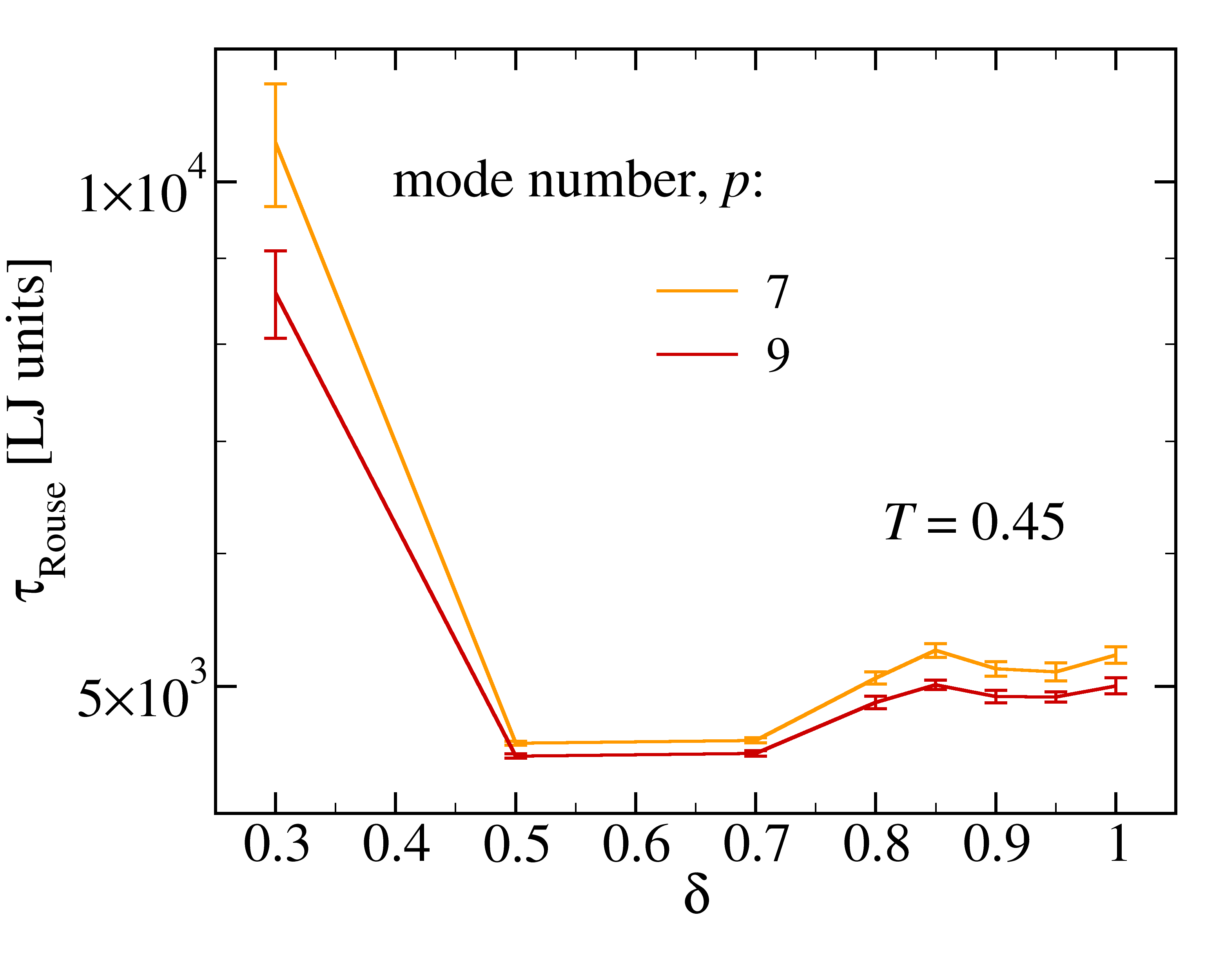}
	
	\caption{(a) Relaxation times of various Rouse modes in a polymer melt containing 20\% spherical additive particles. The horizontal axis is the additive/monomer size ratio. Since variations of $\tau_{\mathrm{Rouse}}$ are quite small for the mode numbers $p=7$ and $p=9$, these data are repeated in panel (b). This panel thus serves to better visualize the existence of multiple extrema for these large mode numbers. The panels (c) and (d) show exactly the same type of data but at a different temperature as indicated. Both temperatures belong to the supercooled state of the polymer melt. Error bars give the 86\% confidence interval.}
	\label{fig:Rouse-tau-vs-delta}
\end{figure}

Results on the dependence of relaxation times of Rouse modes on size ratio $\delta$ are shown in Fig.~\ref{fig:Rouse-tau-vs-delta}.  The non-monotonic effect and the existence of multiple extrema are clearly visible in this figure for all Rouse modes investigated. This observation supports the view that connectivity along the chain's backbone has no major effect on the coupling between dynamics of additive molecules and polymer chains. Interestingly, also the size ratios for which different extrema occur, seem to be largely independent of the mode index $p$. This suggests that the phenomenon can be rationalized letting aside polymer specific features. Following this idea, we provide below a study of the same effect in a binary mixture of hard spheres. It is shown that local packing effects are responsible for the non-monotonic effect and the occurrence of multiple extrema.

\section{Binary hard-sphere (HS) mixtures}

The simplest model to capture excluded-volume interactions and specifically the ensuing mixing effects of particles with different sizes, is the binary hard-sphere mixture. We now turn to a discussion of MCT predictions for this system to support the interpretation given above for the dynamics of the polymer/additive system. The overall scenario predicted by MCT for the binary HS model has been discussed at length \cite{Goetze2003,Voigtmann2011}, so we focus on the specific aspects that are relevant for the present discussion.

In the binary HS system, the role of temperature is replaced by that of the overall packing fraction $\varphi$; in the monodisperse HS system with the Percus-Yevick structure factor, MCT predicts a glass-transition point $\varphi_c\approx0.5159$, thus we choose packing fractions $\varphi=0.514$ and $0.515$ as proxies for the temperatures close to $T_c$ that have been studied in the simulation. Essentially, as one enters the asymptotic regime of small $|\varphi-\varphi_c|$, the effects to be discussed below do not change qualitatively, but become more pronounced as one approaches $\varphi_c$.
For simplicity, we
keep the overall packing fraction of the binary HS system constant as well
as the number concentration of small particles, and vary the size ratio
$\delta$ as the single control parameter.

\begin{figure}
    \centering
    (a)\includegraphics[width=0.45\textwidth]{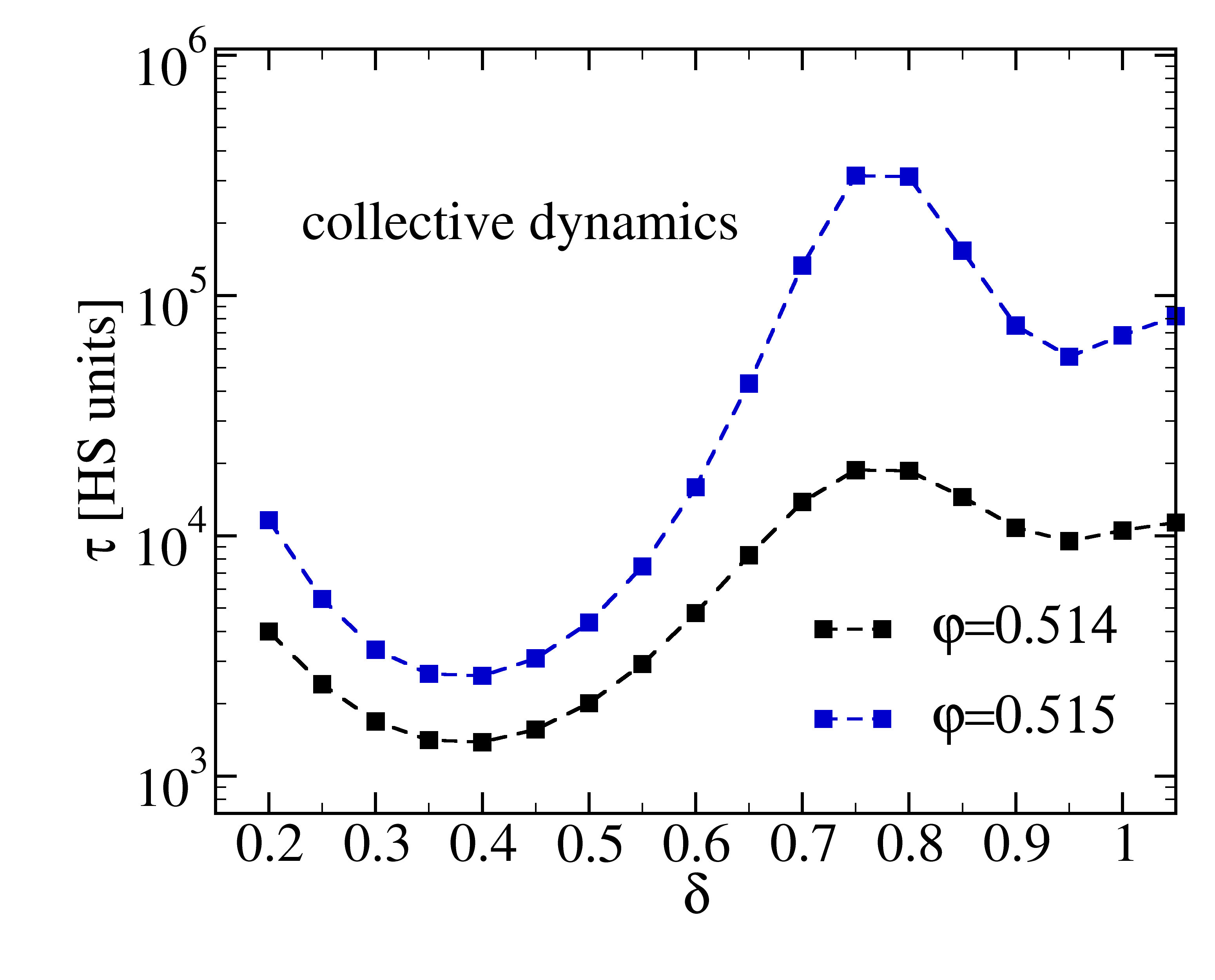}
    (b)\includegraphics[width=0.45\textwidth]{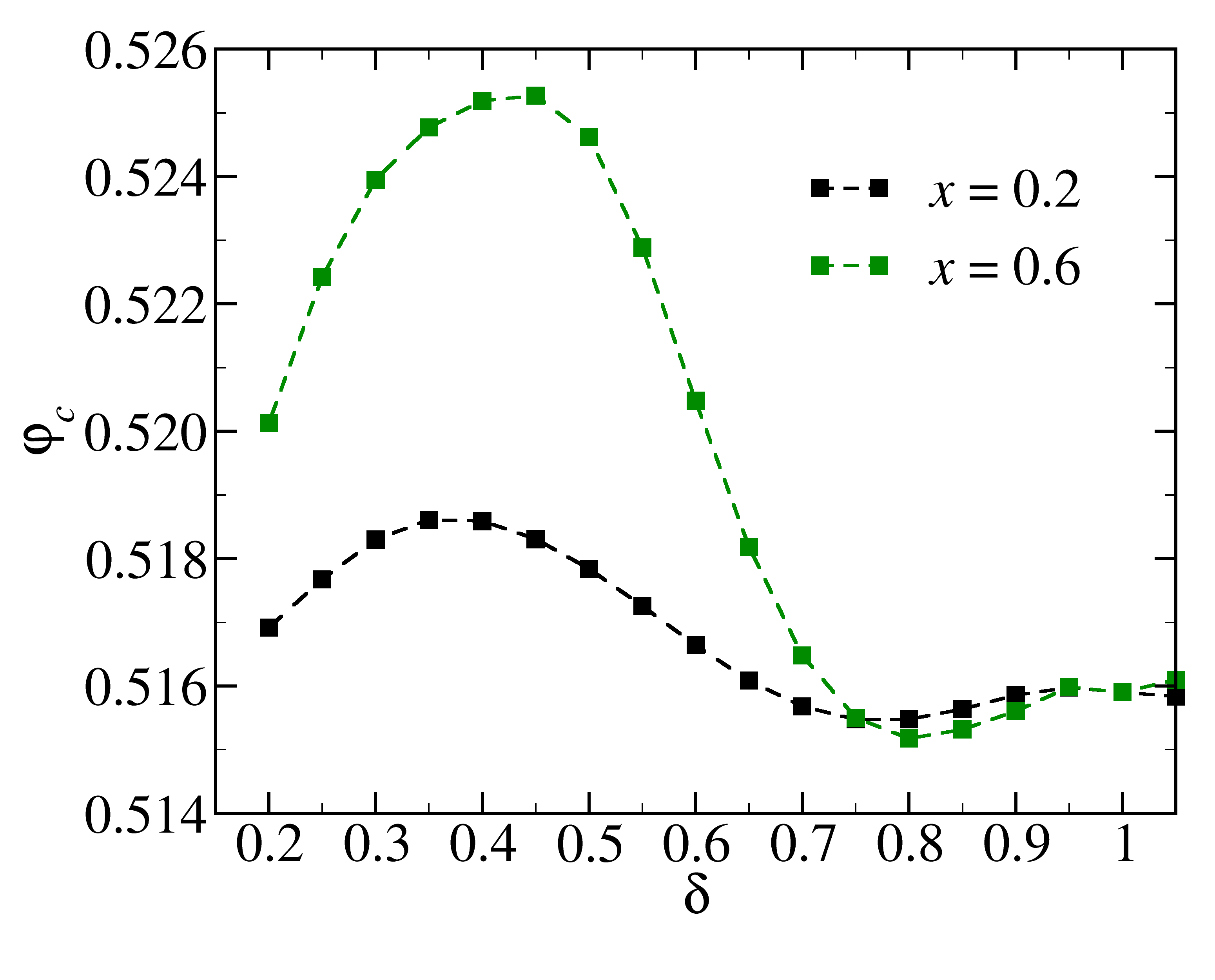}
    \caption{(a) Relaxation times obtained from the decay of collective density auto correlation function of the large particles in a binary HS mixture, obtained from MCT calculations for two packing fractions ($\varphi=0.514$ and $\varphi=0.515$). A non-monotonic trend with two minima and one maximum can be seen in the range $\delta\leq1.0$. Panel (b) shows the corresponding variation of the critical glass-transition packing fraction $\varphi_c(\delta)$ at fixed concentration $x=0.2$.}
    \label{fig:MCT_tau}
\end{figure}

To quantify the change in relaxation of the collective dynamics, we show in Fig.~\ref{fig:MCT_tau}a the structural relaxation times $\tau$ extracted from the density correlation function of the large particles at a wave number $q=7$, corresponding to the main peak of the static structure factor in the monodisperse system. As is evident already for $\varphi=0.514$ and more pronounced for $\varphi=0.515$, this relaxation time displays a maximum around $\delta=0.8$, and a minimum around $\delta=0.4$, for the chosen composition. There is an additional slight maximum around $\delta=1.05$. This is in qualitative agreement with the non-monotonic variation discussed in Fig.~\ref{fig:RelaxTime_MSD}, where a maximum around $\delta=0.85$ and
a minimum around $\delta=0.6$ are seen; also there, a slight maximum is visible around $\delta=1$. We attribute the small differences in $\delta$ values to the differences in the models: In particular, for the HS mixture that case $\delta=1$ implies that there is no difference between large and small particles, while for the polymer/additive system even at $\delta=1$ there are structural and dynamical differences between the monomers of a polymer and the same-sized additive particle.

Also shown in Fig.~\ref{fig:MCT_tau}b are the critical packing fractions $\varphi_c$ found for the various size ratios. The $\varphi_c$-versus-$\delta$ curve shows a clear maximum around $\delta=0.4$, and a minimum around $\delta=0.8$. These two extrema thus explain the pronounced corresponding minima and maxima of the relaxation times -- as the overall packing fraction of particles is kept constant, one moves further away from, respectively, closer to the glass transition point. These extrema are also robust against changing the small-particle concentration within a certain range, as can be seen by noticing that the $\varphi_c$-versus-$\delta$ curves shown in Fig.~\ref{fig:MCT_tau}b for two different concentrations display
qualitatively the same variation. The additional extrema seen in the relaxation-time curve do not have corresponding extrema in the variation of the glass-transition point. They are thus more subtle features of the asymptotic approach to the glass transition.

\begin{figure}
    \centering
    (a)\includegraphics[width=0.45\textwidth]{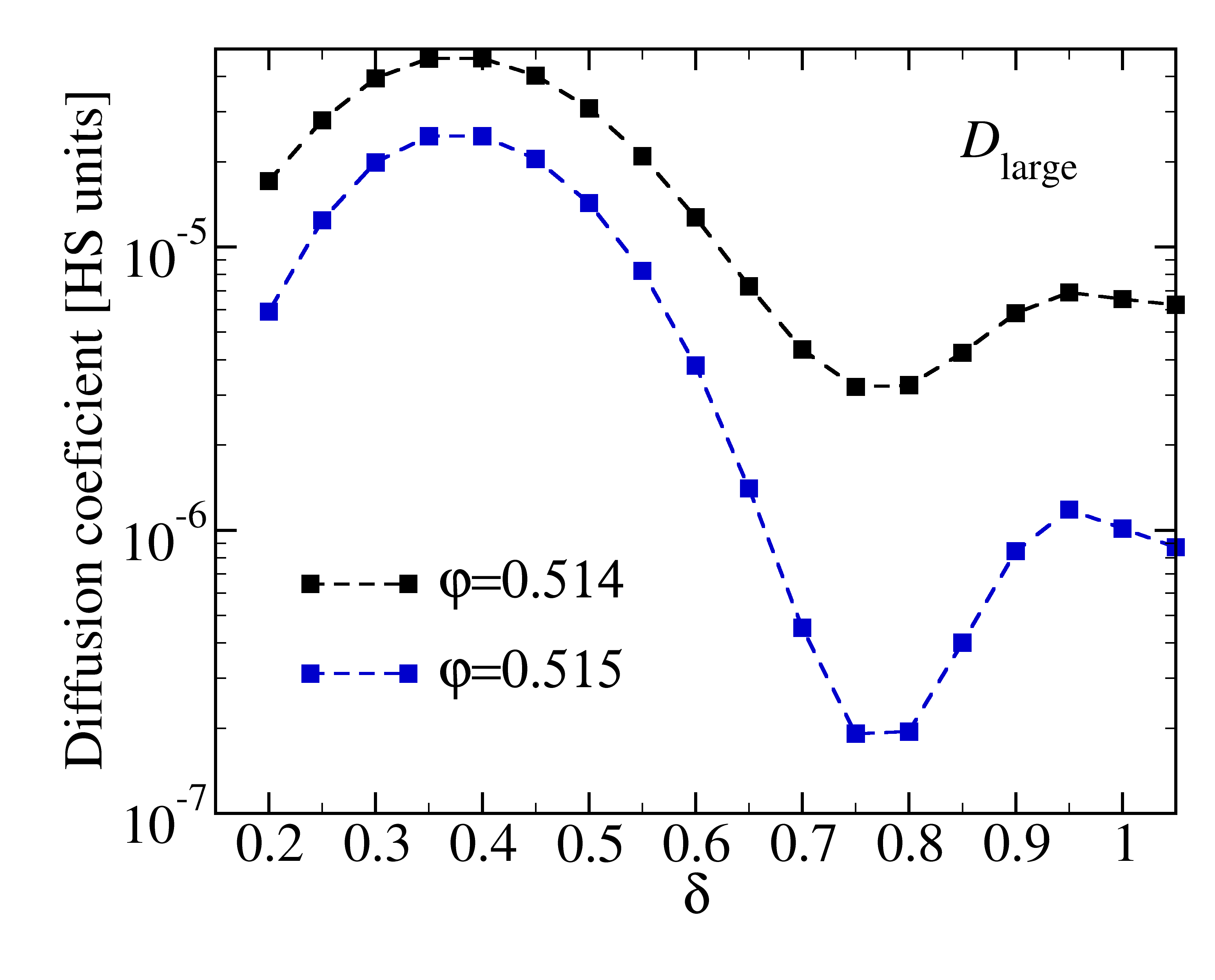}
    (b)\includegraphics[width=0.45\textwidth]{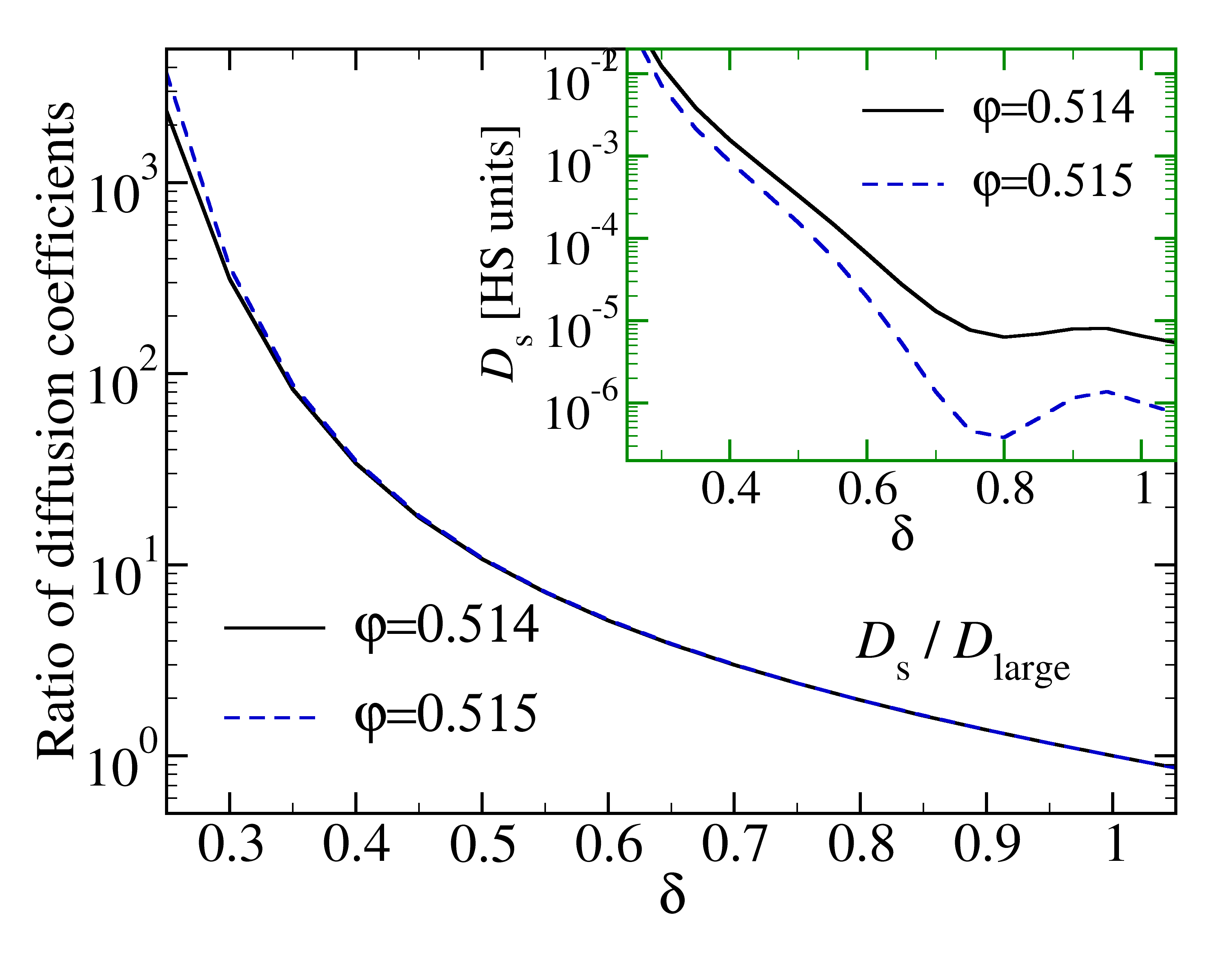}
    \caption{Diffusion coefficient, $\Dl$, of the large particles in a binary HS mixture versus the size disparity for two packing fractions as obtained from MCT. (b) The ratio of the small-particle diffusion coefficients to that of the large particles, $\Ds/\Dl$, showing a steep increase for $\delta \to 0$ in qualitative agreement with the observation for the polymer/additive system. The inset shows variation of small particles' diffusion coefficient, $\Ds$, with the size ratio.}
    \label{fig:MCT_diff}
\end{figure}

The diffusion coefficients predicted by MCT for the binary HS mixture,
Fig.~\ref{fig:MCT_diff}, likewise capture qualitative trends that
are visible in the results for the polymer/additive system,
Fig.~\ref{fig:DiffCoef_T46}. At small $\delta$, a decoupling of the
diffusivity of the smaller particles from the diffusivity of the rest
of the system becomes apparent; this is a precursor to the transition to a partially arrested glass that MCT predicts in this range of size ratios.
As a consequence, the small-particle diffusivity for small $\delta$ has
a weaker dependence on $\varphi$ than the large-particle diffusivity,
because it is not influenced by the vicinity of $\varphi_c$. This effect is qualitatively similar to the observed decoupling in the polymeric system~\cite{Zirdehi2019}. It, in particular, explains
the weak temperature dependence of the additive diffusivity seen in the
simulations for small $\delta$. At the same time, the large-particle
diffusivity shows a maximum around $\delta=0.4$ as expected from the
collective relaxation dynamics, and this maximum qualitatively agrees with the one occurring around $\delta=0.6$ in the polymer/additive system.

At large $\delta$, differences appear that are again due to the simplification made in the binary HS model: here, necessarily the ratio of diffusion coefficients approaches unity as $\delta\to1$; this is not the case in the polymer/additive system, where equal-sized free particles remain faster diffusing than their monomeric counterparts in the polymer chains.
This originates from chain connectivity, which slows down the dynamics of a monomer as compared to a single non-bonded particle.
It is, therefore, quite interesting that, despite this asymmetry, all the qualitative features remain similar between polymer-additive system on the one hand and binary HS mixture on the other hand. This again underlines the dominance of local packing effects in the phenomenon addressed in this work.

\section{Summary and outlook}

In this study, the effect of size disparity on the glass transition is explored for two different systems, namely polymer-additive blends and hard-sphere mixtures. The first system is studied by molecular dynamics simulations and the latter via mode-coupling theory. In the polymeric systems, monomers act as larger components which are connected to each other forming linear chains. The smaller component (additive molecule) is modeled as spherical LJ particles. Relaxation time is found to exhibit multiple extrema upon a variation of the additive/monomer size-ratio at constant additive's number concentration. The trend is robust and shows itself in structural relaxation at various length scales ranging from a monomer diameter to the chain's end-to-end vector. Interestingly, a similar trend is found also in mode coupling theoretical calculations of a binary hard sphere mixture. These observations suggest an interpretation of the additive's effect in terms of local packing effects and the related coupling between additive molecules/minority species and polymer/majority species. Essentially, the small additives are most effective in speeding up the relaxation dynamics when they are sufficiently different in size from the majority species, but also not too small to merely diffuse inside the structure set by the larger particles. This leads to a strong acceleration in relaxation around $\delta\approx0.4$ ($\delta\approx0.6$ in the polymeric system). If the additives are just slightly smaller than the majority species ($\delta\approx0.8$), their smaller size increases free volume, but does not yet disrupt the overall structure, so that in the HS system, a slightly smaller overall packing fraction is sufficient to achieve the same slow relaxation as in the monodisperse system.

The interpretation of the additive effect given here in terms of local packing effects between monomers and additive molecules is also borne out in the application of MCT to polymer melts~\cite{Chong2002,Chong2007}. It would be interesting to see if the inclusion of specific intra-chain interactions in that theory will be able to improve the qualitative agreement that we found here also for large $\delta$.
A further interesting question concerns ways to control this non-monotonic effect via additional mechanisms. One possibility would be to tune interaction energies in order to make the interactions between the two species more repulsive.

\section*{Acknowledgments} The authors acknowledge financial support by the German Research Foundation (DFG) project Nr. VA205/16-2 within the SPP1713.

\section*{References}
\bibliographystyle{prsty_with_title}
\bibliography{literature-2019-10-10}
\end{document}